# Bipolar Anodization Enables the Fabrication of Controlled Arrays of TiO$_2$ Nanotube Gradients

G. Loget,[a] S. So,[a] R. Hahn[a] and P. Schmuki[a,b]*

[a] Department of Materials Science WW-4, LKO, University of Erlangen-Nuremberg, Martensstrasse 7, 91058 Erlangen (Germany)

[b] Department of Chemistry, King Abdulaziz University, Jeddah (Saudi Arabia)

* Corresponding Author. Schmuki@ww.uni-erlangen.de

**Abstract**

**We report here a new concept, the use of bipolar electrochemistry, which allows the rapid and wireless growth of self-assembled TiO$_2$ NT layers that consist of highly defined and controllable gradients in NT length and diameter. The gradient height and slope can be easily tailored with the time of electrolysis and the applied electric field, respectively. As this technique allows obtaining in one run a wide range of self-ordered TiO$_2$ NT dimensions, it provides the basis for rapid screening of TiO$_2$ NT properties. In two examples, we show how these gradient arrays can be used to screen for an optimized photocurrent response from TiO$_2$ NT based devices such as dye-sensitized solar cells.**

# Introduction

Over the past decade, arrays of aligned $TiO_2$ nanotubes ($TiO_2$ NTs) grown on Ti by self-organizing anodization have attracted a huge scientific interest.[1,2] Due to the electronic properties and the biocompatibility of $TiO_2$, these one-dimensional nanostructure arrays are highly promising for applications in the fields of dye-sensitized solar cells (DSSCs),[1,3-4] water splitting,[5] photocatalysis[6] and biomedical devices.[7,8] The performance of the nanotubes in all these applications depends strongly on their length, diameter and ordering. For instance, an optimal length of 7 μm for $TiO_2$ nanotubes has been reported for solar photoelectrochemical water splitting,[9] ranges of 15-30 μm seem optimum for DSSCs,[4] and it has been shown that $TiO_2$ nanotubes with a diameter larger than 50 nm impaired the spreading and adhesion of mesenchymal stem cells on nanotubular surfaces.[7,10] Screening the properties of $TiO_2$ NT arrays as a function of the tubes characteristic dimensions for a given application is consequently of significant importance for the development and the optimization of many important $TiO_2$ NT-based devices. In the majority of the cases, screening requires fabricating and studying series of individual surfaces, each of them comprised of one type of nanotubes with a specific characteristic tube dimension, this approach is time consuming and expensive. Therefore, it would be beneficial to develop a reliable and efficient technique that allows the fabrication of $TiO_2$ NT arrays with adjustable length and diameter gradients for the rapid screening of these factors on a single sample. In the present work we show that such gradients can be generated in a highly controlled manner using bipolar electrochemistry.

Bipolar electrochemistry is a phenomenon which generates redox reactions on the surface of conductive objects without the use of wires.[13] It has recently attracted considerable attention for micro- and nanosciences with applications in the domains of analysis,[14] materials science,[15] as well as for the generation of particle motion.[16] Although bipolar electrochemistry can be conceived as a straightforward fabrication technique of various potential induced surface gradients, only a few papers describe its use for this purpose. Björefors *et al.* used bipolar electrochemistry-triggered desorption of self-assembled monolayers for the formation of molecular gradients on gold,[17] Shannon *et al.* reported the bipolar electrodeposition for the fabrication of Au-Ag and CdS solid-state gradients,[18] and Inagi and Fuchigami used bipolar doping to fabricate gradually-doped conducting polymers.[19] In the present paper, we combine for the first time anodization and bipolar electrochemistry in order to fabricate in a few minutes and in a wireless manner self-assembled $TiO_2$ NT surfaces, which are comprised of NT gradients with tunable length and diameter and we show how these gradients can be used for the rapid screening of the $TiO_2$ NT properties.

**Principle and mechanism**

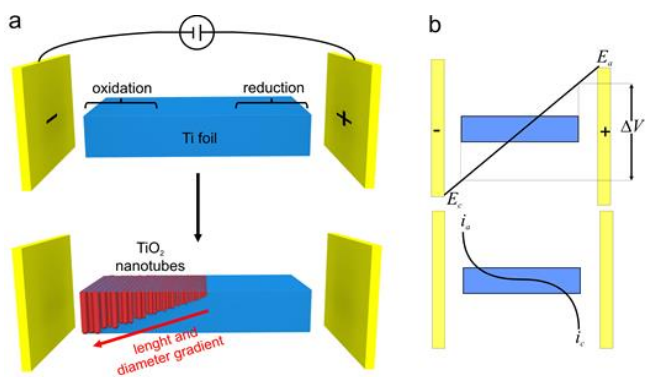

**Fig. 1.** a) Scheme showing the principle of bipolar anodization. The Ti foil becomes a bipolar electrode under a sufficiently high electric field, which leads to the formation of a gradient of TiO$_2$ NTs at its anodic pole. b) Scheme showing the polarization potential distribution (top) and the faradaic current distribution (bottom) on a bipolar electrode.†

Fig. 1 illustrates the principle of bipolar anodization. First, a Ti foil located between two feeder electrodes lies in the electrolyte with no direct electrical connection. When an electric field is applied between the feeder electrodes, namely an anode at a potential $E_a$ and a cathode at a potential $E_c$, a polarization potential $\Delta V$ arises along the surface of the Ti foil.[13] As shown in Fig. 1b, this potential is maximal at the edges of the foil and gradually decreases towards its middle. If the generated polarization potential is high enough, a fraction of the delivered current flows through the Ti foil, inducing redox reactions along the metal conducting path but most pronounced at its extremities, where the polarization potential is maximum.[13] Reduction happens at the cathodic pole of the Ti foil (the extremity which faces the feeder anode), together with oxidation at the anodic pole (extremity which faces the feeder cathode). Under such conditions the Ti foil behaves at the same time as an anode and a cathode, that is a bipolar electrode. Since the driving force of the electrochemical reaction, $\Delta V$, gradually decreases from the edge of the reactive pole towards the middle of the foil, this mechanism, performed under appropriate anodization conditions, should allow to grow TiO$_2$ NTs with a length and diameter gradient on the anodic pole.

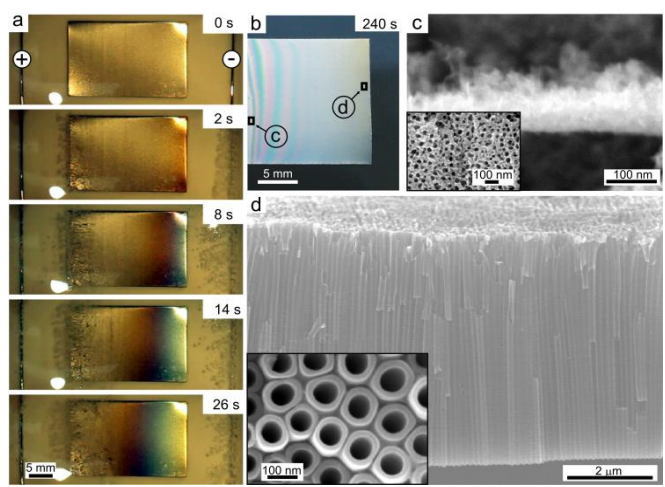

**Fig. 2.** a) Series of photographs showing the first 26 s of a bipolar anodization experiment. b) Photograph of the anodic pole of the Ti foil after 240 s of bipolar anodization. c) SEM cross-section of the oxide layer located at spot c in Fig.2b. Inset: SEM top view of the layer. d) SEM cross-section of layer located at spot d in Fig. 2b. Inset: SEM picture showing the sections of TiO$_2$ NTs.

Fig. 2 and video 1 (provided in supplementary information) show a proof-of-principle experiment in which a 2.5 cm-long Ti foil immersed in an ethylene glycol-based electrolyte containing lactic acid (recently developed in our lab for the ultrafast growth of ordered TiO$_2$ NTs)[20] was polarized by a 37 V.cm$^{-1}$ electric field. At 2 s, the color of the anodic pole became brownish due to the formation of a titanium oxide layer interfering with some wavelengths of the visible light.[21] Simultaneously bubble evolution occurred at the opposite edge (the cathodic pole) of the foil, caused by the reduction of the electrolyte. Over time the colored front extended towards the cathodic pole, together with the emergence of a linear interference color gradient (from brown to white) caused by the thickness increase of the Ti oxide layer over the surface (thickness = $f_{ox} U_x$ with $f_{ox}$ = oxide growth rate and $U_x$ is the local effective potential at a given surface location) Fig. 2b shows the anodic pole of the Ti foil after 4 min of anodization. The oxide layer, mostly greyish, covered more than 60 % of the Ti foil surface. Scanning electron microscope (SEM) observations were performed at the two extremities of the anodic pole. A porous oxide layer (Fig. 2c) with a thickness of about 100 nm and non-organized nanopores (Fig. 2c, inset), was observed at the edge which underwent the weakest polarization potential. In contrast, as shown in Fig. 2d, the opposite edge, which underwent the highest polarization potential, carried a 5.5 µm-thick layer composed of self-organized $TiO_2$ NTs. The presence of such $TiO_2$ NT layer over the anodic pole suggest that the anodic reactions happening on this area during polarization involve the oxidation of the metal Ti surface to form $TiO_2$ and soluble $Ti^{4+}$ complexes with fluoride, as it is the case for conventional electrochemical formation of $TiO_2$ NTs.[1] This proof-of-principle experiment shows that bipolar anodization can be used for the fabrication of oxide layers with different thicknesses on one Ti foil, but more importantly also layers of self-organized $TiO_2$ NT arrays can be formed. We demonstrate in the following examples that the concept of bipolar anodization can be used for the straightforward, rapid and reliable fabrication of $TiO_2$ NT dimensional gradients.

**Control over the distribution of the nanotube lengths and diameters**

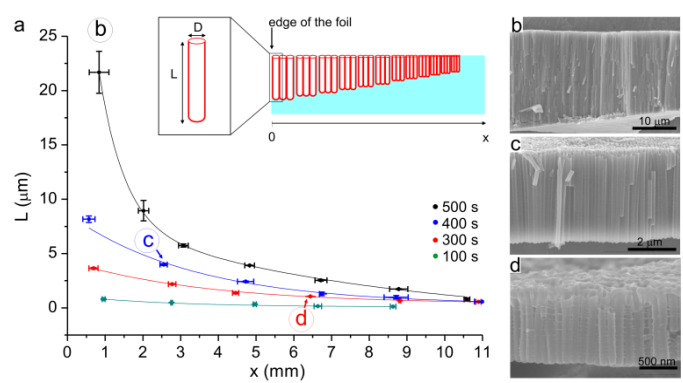

**Fig. 3.** a) Curves showing the evolution of L as a function of x for different bipolar anodization times using a 55 V.cm$^{-1}$ electric field. Inset: scheme of a $TiO_2$ NT gradient varying along the x axis and characteristic dimensions of a single tube, namely length L and diameter D. b), c) and d) SEM pictures showing cross-sections of the oxide layers respectively obtained at the b, c and d points of Fig. 3a.

The inset of Fig. 3a defines the position x on the Ti foil as well as the characteristic dimensions of single NTs, length L and diameter D. Fig. 3b shows gradients obtained after imposing an electric field of E = 55 V.cm$^{-1}$ on 1.8 cm-long Ti foils for different times (a detailed description of the electrochemical cell is provided in Fig. S1). First, it is clear that in every case the longest $TiO_2$ NTs were obtained at the edge of the anodic pole and their length decreased in the direction of the other edge of the foil. The length decrease is found of an exponential nature which suggests a correlation between L and the current distribution and that the anodization reaction is not limited by mass transfer because the current distribution over a bipolar electrode (see Fig. 1b) follows the exponential Butler-Volmer relationship in the case of electron-transfer limited reaction.[14,22,23] A very useful finding is that the height of the gradient can be easily tuned by the charge passed through the bipolar electrode, thus, simply by the time of the electrolysis. As shown in these graphs, the longest tubes (L ≈22 µm) were obtained for an electrolysis time of 500 s. At this position, the L gradient can easily be seen in the cross-section SEM of Fig. 3c. The smallest measured tubes, obtained with an electrolysis time of 100 s, were shorter than 500 nm. These results show that $TiO_2$ NT length gradients can be easily fabricated and tuned by bipolar anodization.

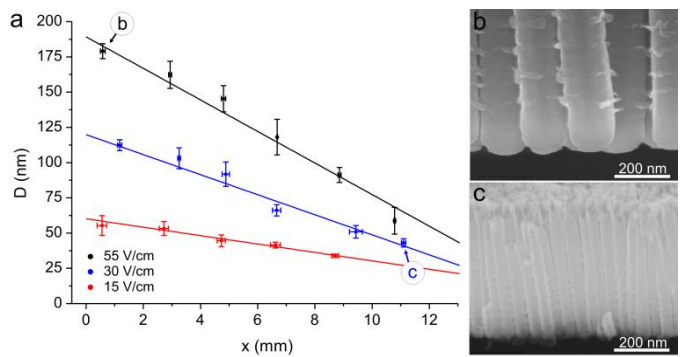

**Fig. 4.** a) Curves showing the evolution of diameter D as a function of x for different applied electric fields. b) and c) SEM pictures showing the TiO$_2$ NTs respectively obtained at the b and c points of Fig. 4a.

The tube diameters were not significantly affected by the time of electrolysis (see Fig. S6) but directly by the applied electric field, as shown in Fig. 4. The bipolar anodizations performed for this figure were performed at three different electric fields: 55, 30 and 15 V.cm$^{-1}$. As the growth rate decreased with decreasing electric field values, the electrolysis was performed during a longer time (t = 4, 14 and 75 min, respectively). The D gradient curves, shown in Fig. 4, demonstrate that the evolution of D as a function of the position x is linear over the anodic pole, which suggest that it is controlled by the variation of the polarization potential over the anodic pole (see Fig. 1b).[13] This is consistent with the fact that the D gradient is adjustable with the applied electric field because the polarization potential is proportional to the imposed electric field.[13] The largest tubes, shown in Fig.4b, had a diameter of 180 nm and were obtained for E = 55 V.cm$^{-1}$. In this case the ratio $D_{max}/D_{min}$ = 3 with a slope of 11.2 nm.mm$^{-1}$. The narrowest measured tubes had a diameter smaller than 35 nm and were obtained with E = 15 V.cm$^{-1}$ (slope = 3.0 nm.mm$^{-1}$). These results illustrate that TiO$_2$ NTs also with a gradient in diameter can be easily fabricated and tailored using bipolar anodization.

**Photocurrent screening**

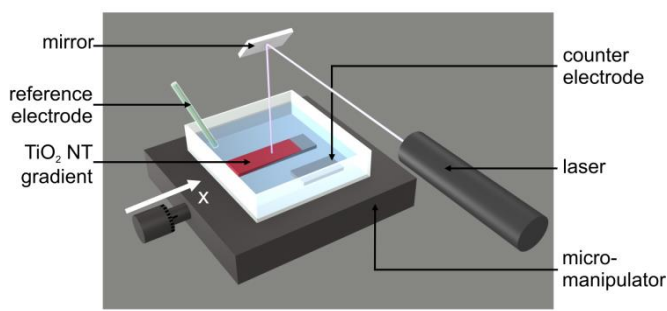

**Fig. 5.** Scheme of the set-up used for photocurrent screening.

After having shown the fabrication of TiO$_2$ NT gradients, we now demonstrate two examples how they can be used for the fast screening of the NT properties. Motivated by the current scientific discussion in the literature about the optimal characteristic tube dimensions that provide a maximum photoresponse e.g. in photocatalysis or DSSCs,[4,6,9] we screened our gradients in a photoelectrochemical set-up, as illustrated in Fig. 5.

The gradients used in these experiments were fabricated by bipolar anodization on 3.8 x 1 cm Ti foils (see supporting information for experimental details). As shown in Fig. 6b, those surfaces comprised 2.4 cm-long TiO$_2$ NT gradients with a ratio between the maximum and minimum length values L$_{max}$/L$_{min}$ ≈100. The gradients were annealed to anatase form (see Fig. S4 for X-ray diffraction spectrum) and placed in the photoelectrochemical cell (experimental details are given in supporting information). This cell consisted of a Ag/AgCl reference electrode, a Pt counter electrode and the TiO$_2$ NT gradient foil, used as a working electrode. The cell was filled with the electrolyte and a potentiostat was used for applying a potential to the tube gradient electrode (+500 mV vs Ag/AgCl) and for recording currents. A UV light spot, with a diameter of 1.5 mm (FWHM of the Gaussian intensity distribution), generated by a HeCd laser (325 nm) was used for illuminating the gradient locally. A micromanipulator, mounted below the cell, was used for moving the cell and thus scanning the gradient along its x axis. A dark current I$_d$ ≈2 µA was measured and the transient currents were recorded while imposing 10 s pulses with the laser for different x values, as shown in the inset of Fig. 6a.

Fig. 6a shows the photocurrent profile along the gradient. From this curve it is clear that the minimum photocurrents (with I$_{min}$ = 0.3 mA) were obtained for lower x value, the values then increased, to a value I$_{max}$ = 1.53 mA for x = 14 mm and decreased down to a value of 1.06 mA for increasing x values. The plots of the photocurrent as a function of the TiO2 NT length and diameter, shown in Fig. 6c, reveal that the highest photocurrent were obtained for L = 6 µm and D = 108 nm. These results are in very good agreement with previous published works.[9] It is also worth mentioning that the transient photocurrent curves obtained with the longest tubes exhibited a decay after illumination (Inset of Figure 6a), this can be attributed to an increased recombination over longer transport pathways. The experiment demonstrates the feasibility to use these gradients for the fast screening of tube properties over a wide range of dimensions.

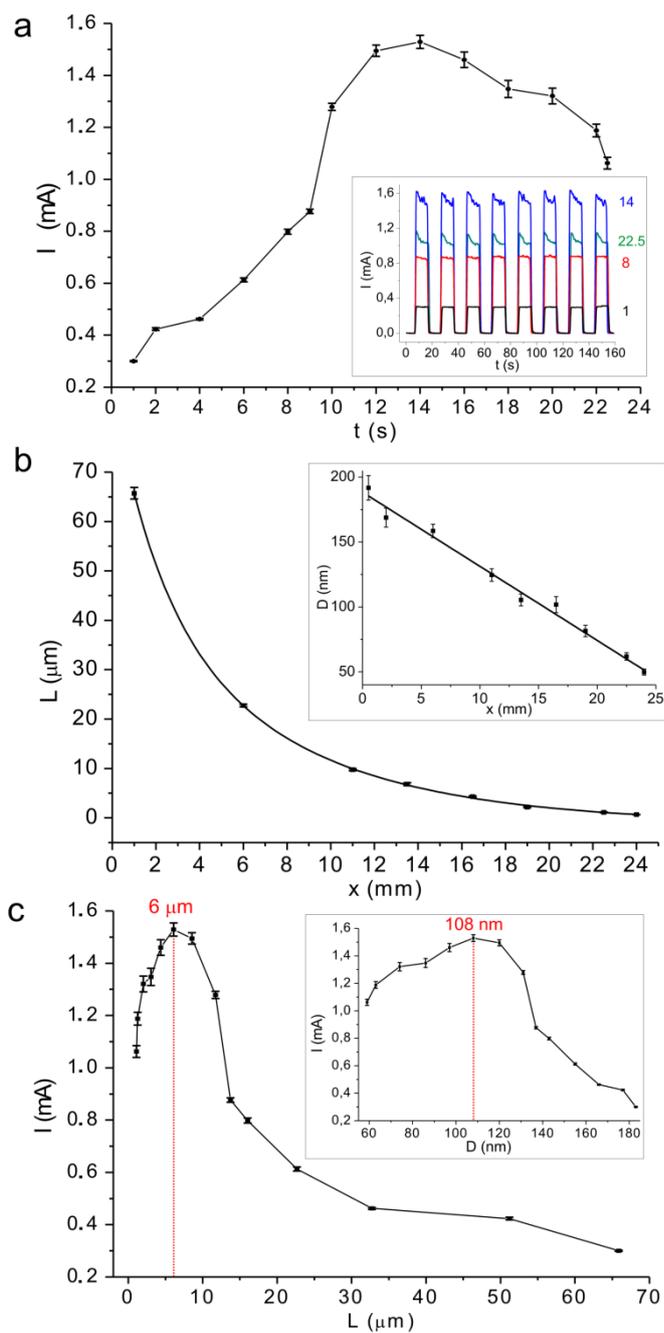

**Fig. 6.** a) Curve showing the photocurrent evolution as a function of x. Inset: transient photocurrent curves for 4 different positions on the gradient (x = 1, 7, 14 and 22.5 mm). b) Length distribution over the TiO$_2$ NT gradient. Inset: diameter distribution over the TiO$_2$ NT gradient. c) Curve showing the photocurrent evolution as a function of L. Inset: Curve showing the photocurrent evolution as a function of D.

**Determination of the optimal dimensions of dye-sensitized TiO$_2$ NTs**

We now show a second direct application of our gradients, that can be used for optimizing TiO$_2$ NT based DSSCs. For these experiments, we used a gradient made similarly to the previously described one. After annealing, the gradient was sensitized with a ruthenium dye and the screening was performed with the set-up similar to the one previously described in Fig. 5. In this case a laser of λ = 473 nm and an acetonitrile-based electrolyte containing the iodide/tri-iodide redox couple were used (see supplementary information, section 1-3 for experimental details). In order to avoid water contamination that could provoke dye desorption, a Pt electrode was used as a pseudo-reference electrode and +500 mV were imposed to the gradient. A dark current $I_d$ ≈7 μA was measured and the current under light irradiation was measured for different x values after 30 s of equilibrium. The photocurrent values as a function of x are shown in Fig. 7a. The maximum photocurrent $I_{max}$ = 70 μA was found for x = 4 mm. The TiO$_2$ NTs present at this location were observed at the SEM and are shown in Fig. 7b. The length of these tubes at x = 4 mm is L = 30 μm with a measured diameter of D = 164 ± 9 nm, which is in very good agreement with Fig. 6b. These dimensions are in the same range than the ones reported in recent work discussing the optimal tube length for DSSCs.[4] The fabrication and the screening experiment was performed in a few hours whereas it would have taken much longer time using classical anodization procedures, probably several days. These results demonstrate clearly how these gradients and the described screening methods can be beneficial for rapidly optimizing photoelectrochemical devices.

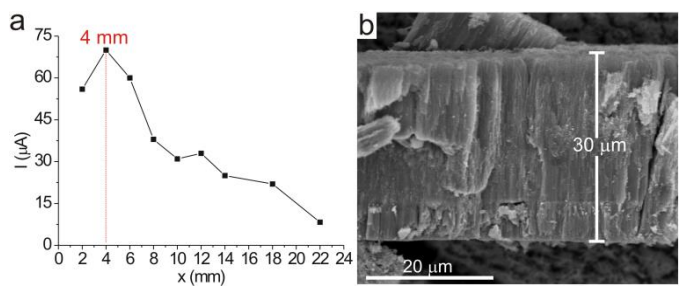

**Fig. 7.** a) Curve showing the photocurrent evolution as a function of x. b) SEM picture showing a cross-section of the oxide layer at x = 4 mm.

**Conclusion**

In summary, we report a new concept of bipolar anodization which allows the rapid wireless fabrication of self-organized TiO$_2$ NTs layers with a gradient in length and in diameter. The length and the diameter of the tubes are respectively correlated with the current and the potential distribution over the bipolar electrode. This makes the height of these gradients easily tunable using simple experimental parameters, such as the time of electrolysis and the applied electric field. The highest ratio between the maximum and minimum measured dimensions was about 100 and 4 for length and diameter respectively, which are the highest dimensional variations reported so far for TiO$_2$ NT gradients.[11-12] The application of these gradients for the fast scanning of the tube properties was demonstrated and allowed determining the optimal tube characteristics for photocurrent generation under UV light (L = 6 μm and D = 110 nm) and for dye sensitized tubes under visible light (L = 30 μm and D = 160 nm). The fabrication and the screening of those gradients could be performed in only a few hours whereas such a broad screening of tube properties performed with conventional anodization techniques would have taken at least several days. This technique thus opens up a path to rapid screening of self-ordered TiO$_2$ NT properties as a function of individual tube dimensions. For example, local photocurrents under different light sources or simulated sunlight or maximum biocompatibility can be screened straightforward; therefore the future optimization of TiO$_2$ NT-based devices in applications such as photovoltaic cells,[1] photoelectrochemical water splitting,[5] and biomedical devices[7] can be drastically faster.

**Acknowledgements**

This work was supported by a post-doctoral research grant from the Alexander von Humboldt Foundation. ERC and DFG are acknowledged for support. Dr. Altomare, Dr. Lee and Prof. Virtanen are acknowledged for valuable discussions.

**Notes and references**

†The potential drops at the solid/liquid interfaces and the mass-transfer effects are considered negligible in these schemes. For more information, see refs. 14 and 22.

Electronic Supplementary Information (ESI) available: video 1, experimental section, characterization of the annealed gradients, screening of the experimental conditions and effect of the electrolysis time on the tube diameter.

See DOI: 10.1039/b000000x/

# Supporting Information

Bipolar Anodization Enables the Fabrication of Controlled Arrays of TiO$_2$ Nanotubes Gradient

Gabriel Loget, Seulgi So, Robert Hahn and Patrik Schmuki*

Table of Contents

# 1- Experimental section

## 1-1- Chemicals and Materials

Chemicals and titanium foils (0.125 mm thick, >99.6 % purity, Advent, England) were used as received. Ethylene glycol (>99.5 %) and DL-lactic acid (90 %) were purchased from Fluka, ethanol (99.5 %) and acetone (Reag. Ph. Eur.) were purchased form Merck. Ammonium fluoride (ACS reag. >98 %) was purchased from Sigma Aldrich. The electrolyte was prepared with ultrapure water (resistivity = 17.1 mΩ.cm). For the fabrication of the arrays, Microposit S1813G2 photoresist and Microposit 351 developer were used.

For dye sensitization, annealed nanotube layers were immersed in a 300 mM solution of Ru-based dye (cis-bis(isothiocyanato)-bis(2,2-bipyridyl-4,4-dicarboxylato) ruthenium(II) bis-tetrabutylammonium) (Sigma Aldrich) in a mixture of acetonitrile and tert-butyl alcohol (1:1 v/v) for 1 day. After dye sensitization, the samples were rinsed with acetonitrile to remove non-chemisorbed dye.

## 1-2- Bipolar Anodization and Characterization

The electrolyte used was previously developed by our lab and used for the fast growing of $TiO_2$ nanotubes.[1] It was composed of ammonium fluoride (0.1 M), lactic acid (1.5 M) and 5 wt% water in ethylene glycol. Titanium pieces of 2.5 x 1.6 cm (Figure 2), 2 x 0.5 cm (Figure 3 and 4) or 4 x 1.2 cm (Figure 6 and 7) were cut, degreased by sonication in acetone, ethanol and water and dried under a nitrogen stream. The description of the cell used in the experiments of Figure 3 and 4 is shown in Figure S1a and the one used for the fabrication of the gradient used in Figures 6 and 7 is shown in Figure S1b. Basically, the cells are composed of a plastic beaker in which a Teflon support has been fixed in the bottom. The Ti sample was tapped on a Teflon detachable support with Kapton tape covering 1 mm its edges of the in order to avoid any motion of the sample during the electrolysis as well as electrochemistry on the bottom of the foil. The support was afterwards placed in the fixed support. The electrolyte (≈75 mL and ≈140 mL for the cells showed in Figures S1a and S1b, respectively) was then poured in the cell and the two feeders electrodes (platinum foils, 1.2 x 1.2 cm) were placed at a distance of 1 mm from the Ti foil edges. The experiements performed in the cell shown in Figue S1b were done under stirring. The electric field (E = U/d, with E being the electric field, U the imposed potential and d the distance between the feeder electrodes) was applied for a certain time using a LAB/SM 1300 power source (ET System). After the electrolysis, the movable support was placed in ethanol for few hours in order to remove the fluorides from the nanotubes and to detach the Kapton tape from the surface. The titanium foil was then rinsed with ethanol, water and dried under a nitrogen stream. The electrolysis cell used for the experiments performed in Figure 2 had to be adapted for video recording (performed with a Canon Ixus 210 camera). It consisted therefore of a weighting boat in which two Pt wires (feeder electrodes) have been placed. All the SEM characterizations were performed using a FE 4800 SEM (Hitachi). Photographs of the samples were taken with a Galaxy S3 mobile phone (Samsung).

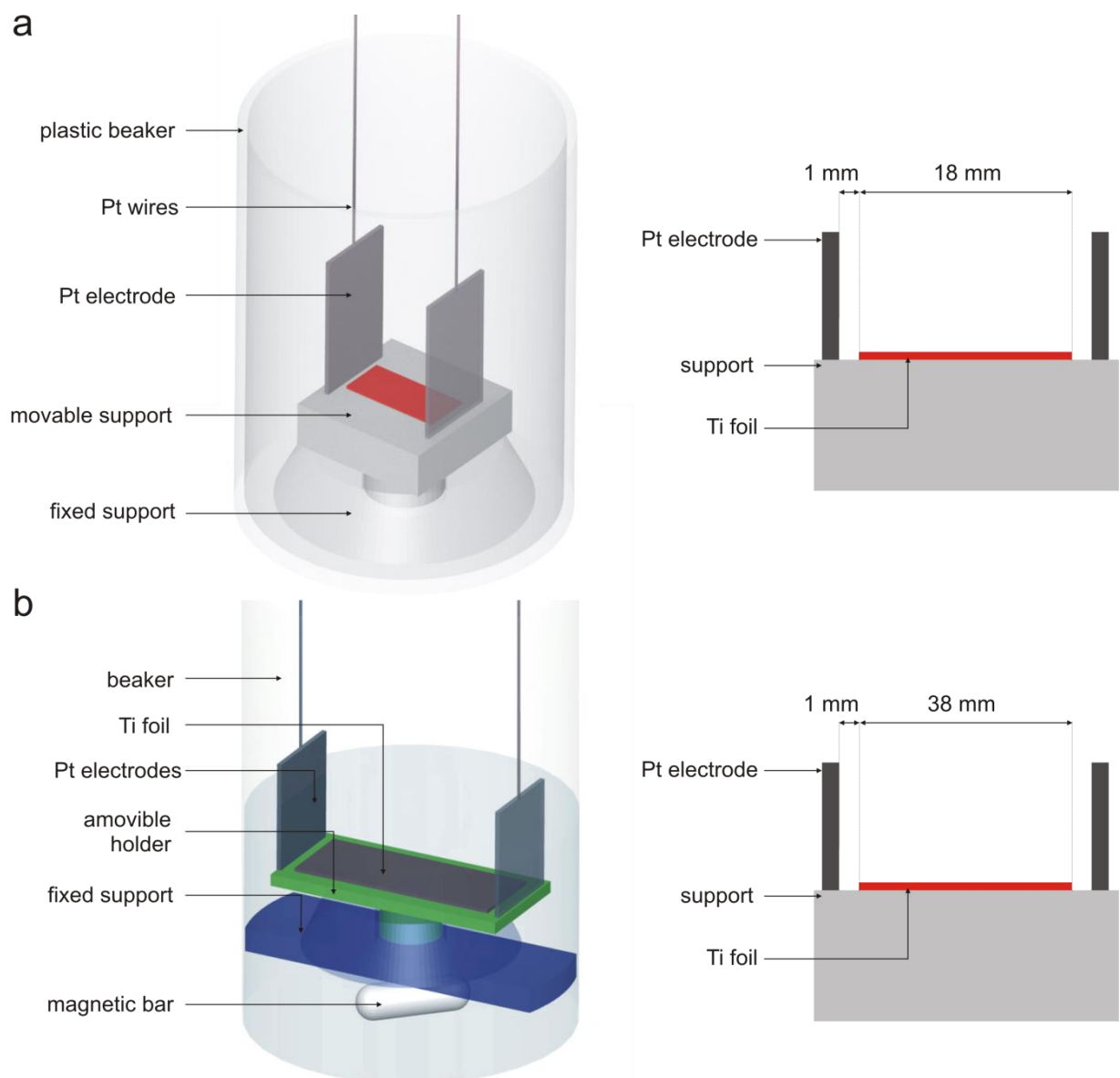

**Figure S1.** Detailed scheme (right side) and cross-section view (left side) of the bipolar electrolysis cells.

1-3- Photocurrent Screening

The non-TiO$_2$ parts on the surface were covered with an insulating polymer layer and the gradient was welded to an insulated Ti wire for making the electrical connection. The reference electrode was a Ag/AgCl electrode (for the experiments of Fig. 6) and a Pt electrode (for the experiments of Fig. 7) and the counter electrode a Pt foil. The counter and the gradient surface were immobilized with tape on the bottom of the cell that was filled with the electrolyte: Na$_2$SO$_4$ (0.1 M) for the experiments shown in Fig. 6 and Iodolyte R50 (Solaronix) diluted with acetonitrile (1:4 v/v) for the experiments shown in Fig. 7. The cell was mounted on a movable stage (Karl Süss) and the laser beam:Kimmon IK Series, 200 mW (for the experiments shown in Fig. 6) or MBL 473, 20 mW (for the experiments shown in Fig. 7) was directed on the middle of the gradient using an aluminum mirror positioned on the top of the photoelectrochemical cell. A potential of +500 mV was applied to the gradient by a potentiostat (Jaissle 1030A) and the transient currents were recorded while imposing cycles of 10 s

"on" and 10 s "off" by opening and closing the laser shutter for the screening shown in Figure 6. For the screening of Figure 7, the currents were recorded after 30 s of equilibrium. The currents were measured at different positions x of the gradient. The lateral photocurrent deviation was measured at 2 positions on the sample and is shown in Figure S2.

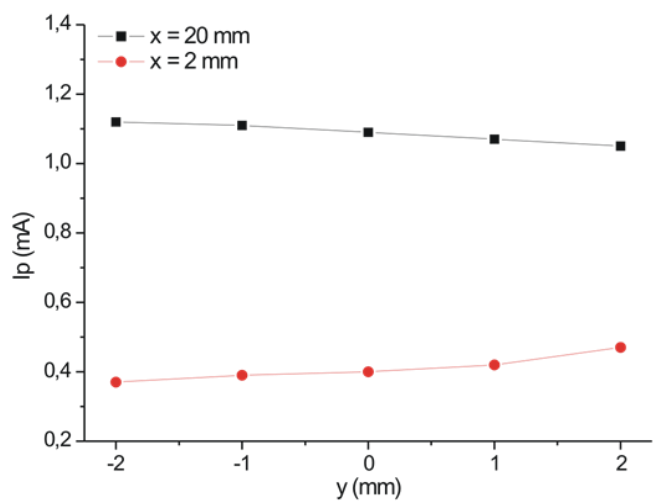

**Figure S2.** a) Curve showing the lateral photocurrent deviation along the y axis at two x positions on the sample.

## 2- Characterization of the Gradient Used for Screening

The gradient used for the screening experiments of Figure 6 was fabricated by bipolar anodization under stirring in the cell shown in Figure S1b. An electric field of 27.5 V.cm$^{-1}$ was applied for 1 h, the delivered current and the temperature of the electrolyte were measured and are shown in Figure S4. The samples were annealed at 450 °C in air with a heating and cooling rate of 30°C min$^{-1}$ over 1 h by using a rapid thermal annealer (Jipelec JetFirst100). X-ray diffraction analysis (XRD, Xpert Philips PMD with a Panalytical X'celerator detector) using graphite monochromized CuKα radiation) was used for determining the crystal structure of the samples. The spectrum is shown in Figure S5.

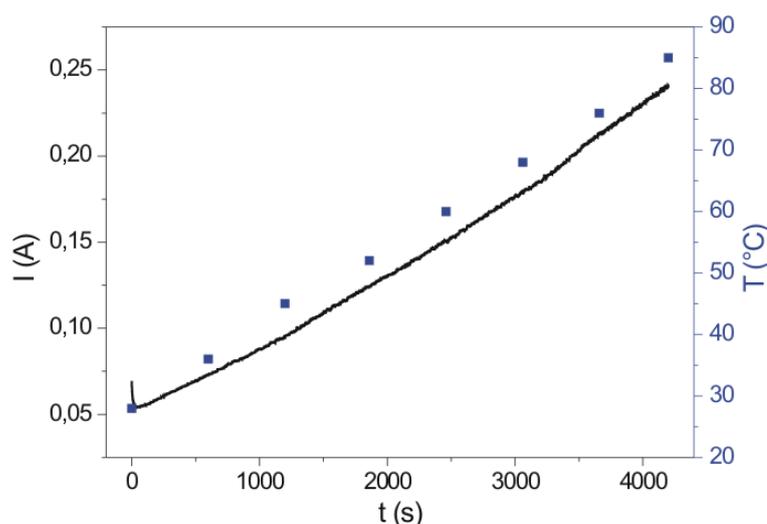

**Figure S3.** Delivered current and temperature of the electrolyte during the bipolar anodization with stirring.

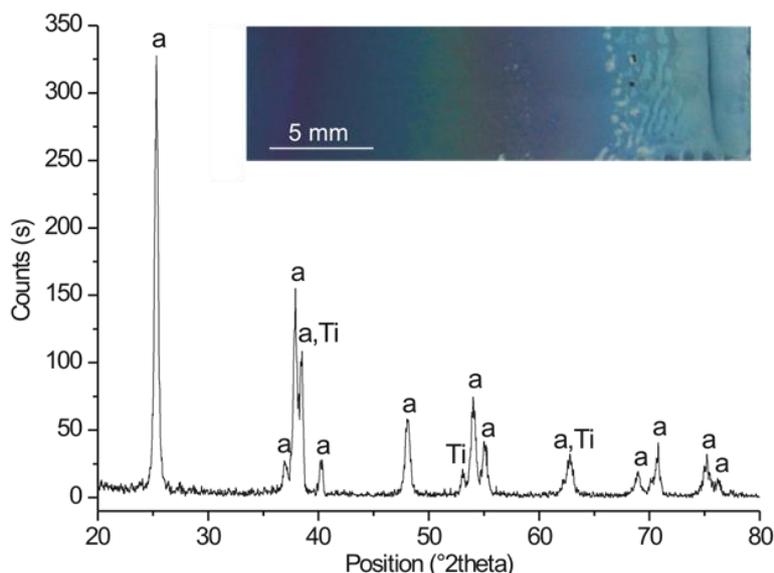

**Figure S4.** XRD spectrum of the annealed gradient. The position of the anatase peaks (a) and titanium (Ti), are indicated. Inset: Photograph of the annealed gradient.

### 3- Screening of Experimental Conditions and Measurements of the Faradaic Current

Different experimental conditions were screened and some of the obtained results are shown in Table S1. Parameters such as stirring, cooling as well as the combination of both were tested. The benefits of stirring as well as cooling is that they allowed maintaining the temperature T and the delivered current $I_d$ at lower values, thus the electrolysis could be performed for longer time before reaching too high T values that may cause boiling of the electrolyte and liberation of hazardous side-products such as gaseous fluoridric acid. Because of the higher charge delivered during the stirred and/or cooled anodization experiments, those conditions allowed obtaining the highest length gradients. Nevertheless, they both strongly limited the growth of the tubes, with decreases of the growth rate by approximately 2 and 3 folds for stirring and cooling, respectively. Furthermore, too long anodization times (t ≥ 30 min) led to the formation of nanograss on the tubes, shown in Figure S5d, caused by the etching of the opening of the tubes by the fluorides.[2] Most of the bipolar anodizations carried out without stirring or cooling led to the classically encountered initiation layer-covered $TiO_2$ NTs, shown in Figure S5a. In the case of bipolar double anodization, a first bipolar anodization was performed, followed by a sonication treatment for removing the tubes, which led to the creation of nanodimples on the surface. When a second bipolar anodization was performed, the tube growth was directed by the dimples, which led to the formation of an initiation layer with holes fitted with the opening of the tubes, shown in Fig S5c. It is worth mentioning that opened $TiO_2$ NTs with a maximum value $L_{max}$ = 32 μm, shown in Figure S5b, were obtained using an electric field value of 65 V.cm$^{-1}$ and a time of 240 s, this is explained by the increase of the etching rate of the initiation layer by the fluoride at this electric field value with respect to the experiment performed at 55 V.cm$^{-1}$.

**Table S1.** Study of different experimental conditions of bipolar anodization and characteristic of the obtained tubes: electric field E, time t, delivered current $I_d$ range, temperature T, maximum length $L_{max}$ and the main aspect of the tube opening. Four SEM pictures of characteristic tube openings are provided in Figure S5.

| sample # | electrolysis type | E (V.cm$^{-1}$) | t (s) | $I_d$ range (mA) | T range (°C) | $L_{max}$ (µm) | main aspect of the tube openings | SEM picture of tube openings |
|---|---|---|---|---|---|---|---|---|
| Ti32 | no stirring | 55 | 400 | 70-210 | 27-56 | 8 | initiation layer | S5a |
| Ti39 | no stirring | 65 | 240 | 90-320 | 28-58 | 32 | opened tubes | S5b |
| Ti37 | no stirring-double anodization | 55 | 240 | 60-240 | 26-65 | 3 | initiation layer | - |
| Ti35 | no stirring-double anodization | 65 | 300 | 90-410 | 27-65 | 11 | initiation layer with better fitted openings | S5c |
| Ti26 | No stirring+cooling | 55 | 1320 | 52-300 | 8-50 | 10 | initiation layer | - |
| Ti19 | stirring | 55 | 390 | 60-90 | - | 3,5 | initiation layer | - |
| Ti22 | stirring | 55 | 1800 | 90-430 | 28-93 | 58 | nanograss | S5d |
| Ti25 | stirring-double anodization | 55 | 500 | 100-275 | - | >4 | initiation layer with better fitted openings | - |
| Ti24 | stirring+cooling | 55 | 7200 | 33-93 | 8-34 | >20 | nanograss | - |
| Ti27 | stirring+cooling | 75 | 2700 | 65-370 | 9-60 | >24 | nanograss | - |

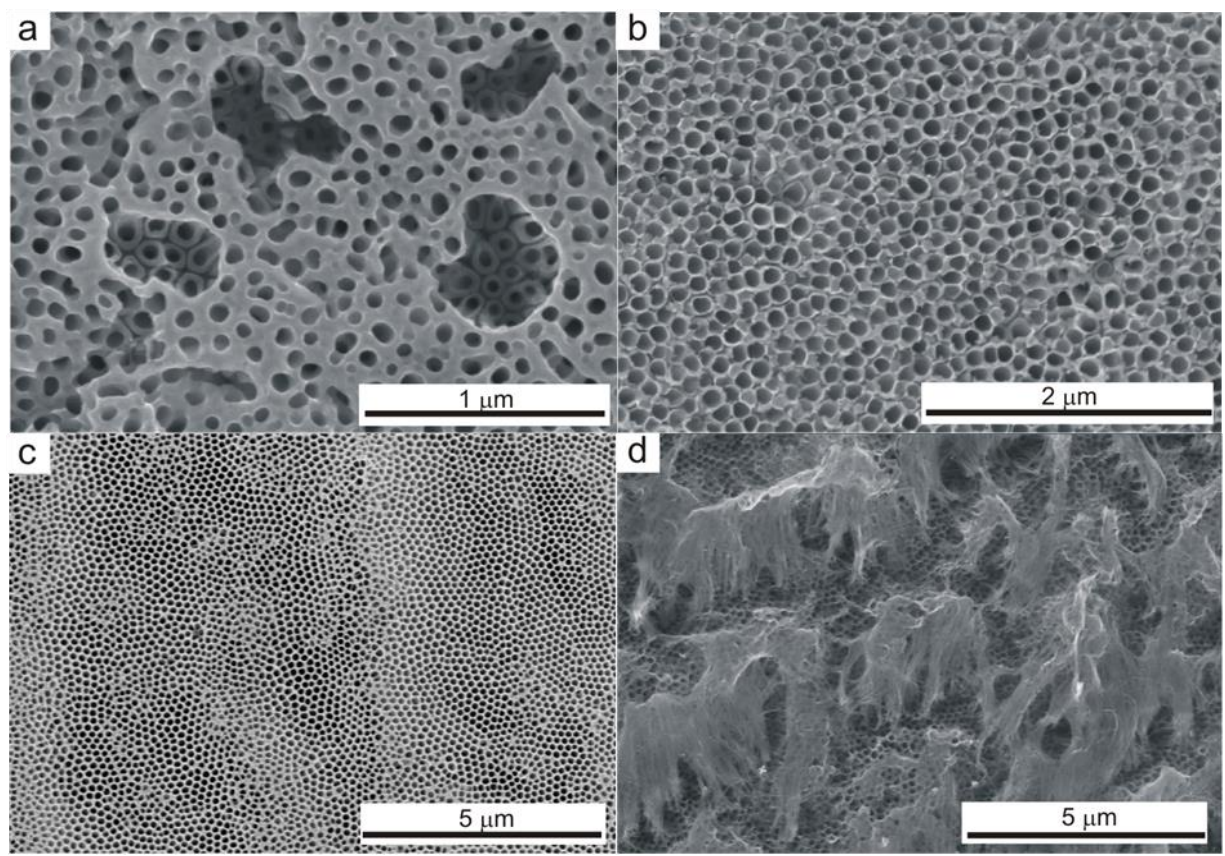

**Figure S5.** SEM top views of TiO$_2$ NT openings obtained by bipolar anodization. a) Tubes covered with an initiation layer (sample Ti32). b) Opened tubes (sample Ti39). c) Tubes covered with an initiation layer having better-fitted openings, obtained by double anodization (sample Ti35). d) Tubes covered with nanograss (sample Ti22).

## 4-Effect of the Electrolysis Time on the Tube Diameter

Figure S6 shows the values of the tube diameters as a function of x for samples obtained with different times of electrolysis using an electric field of 55 V.cm$^{-1}$.

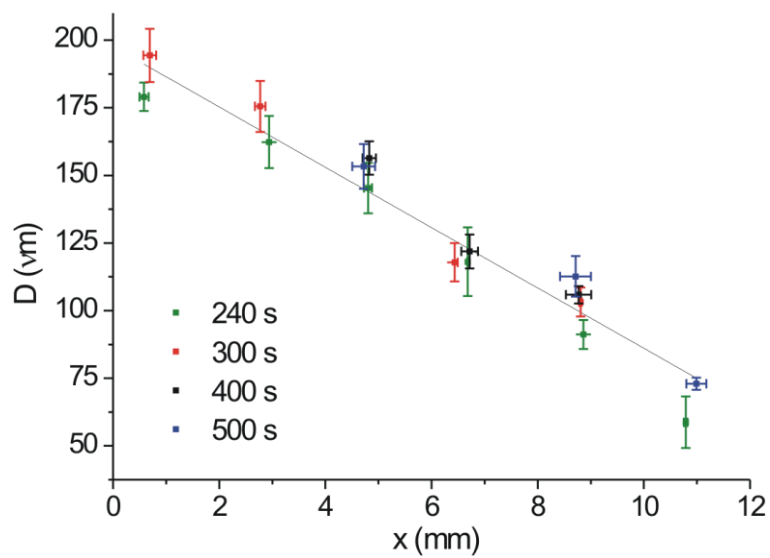

**Figure S6.** Curve showing the values of tube diameters D as a function of x for samples obtained with different times of electrolysis using an electric field of 55 V.cm$^{-1}$.

## 5-References